\documentclass[pra,twocolumn,superscriptaddress,showpacs]{revtex4}
\usepackage{amsmath, amsfonts, amsthm,bbm}
\usepackage[utf8]{inputenc}
\usepackage{natbib} 
\usepackage{graphicx}
\usepackage{color}
\newcommand{\unibasengl}{Department of Physics, University of Basel, Klingelbergstr. 82, 4056 Basel, Switzerland}

\def\fbkt#1#2#3{\mathinner{\langle{#1}\,\lvert\,#2\,\lvert\,#3\rangle}} % Full operator scalar-product
\def\ket#1{\mathinner{\lvert{#1}\rangle}}% Vector
\def\bra#1{\mathinner\langle{#1}|}% Dual Vector
\newcommand{\erw}[1]{\langle {#1} \rangle}% Expectationvalue
\newcommand{\abs}[1]{\lvert#1\rvert}% modulus
\newcommand{\sgn}{\operatorname{sgn}} % signum function
\newcommand{\proj}[1]{\ket{#1}\bra{\!#1}}% Projecting a single variable

\begin{document}
%% Frontmatter
\title{A single fermion in a Bose Josephson Junction}
\author{M.~Rinck}\email{Maximilian.Rinck@unibas.ch}
\author{C.~Bruder}
\affiliation{\unibasengl}
\date{\today}
\begin{abstract}

  We consider the tunneling properties of a single fermionic impurity immersed in a Bose--Einstein condensate in a
  double-well potential. For strong boson--fermion interaction, we show the existence of a tunnel resonance where a
  large number of bosons and the fermion tunnel simultaneously. We give analytical expressions for the lineshape of the
  resonance using degenerate Brillouin--Wigner theory. We finally compute the time-dependent dynamics of the mixture.
  Using the fermionic tunnel resonances as beam splitter for wave-functions, we construct a Mach--Zehnder interferometer
  that allows complete population transfer from one well to the other by tilting the double-well potential and only
  taking into account the fermion's tunnel properties.

\end{abstract} 
%% 67.85.Hj BEC
%% 03.75.Lm BEC-tunneling
%% 67.85.Pq Bose-Fermi mixtures
%% 05.30.Jp Boson Systems
%% 
\maketitle
%% Body

\section{Introduction} %% <<<1

Bose--Einstein condensates (BEC) have become a valuable resource in today's research on many-body physics \cite{Bloch08}.
Loaded into an optical lattice the low-energy regime realizes a Bose--Hubbard Hamiltonian \cite{Jaksch98} whose
parameters can be tuned over a wide range by adjusting the optical lattice or engineering particle interactions via
Feshbach resonances. 

A simpler, however, not less interesting variant of BEC in an optical lattice is obtained, when in the Mott-Insulator
regime, the lattice is modulated by an additional laser beam to create local double-well potentials at each lattice
site. When tunneling between the local double-well potentials is negligible compared to tunneling inside the double-well
potential, the system is well-described by a \textit{two-site} Hubbard Hamiltonian. BEC in double-well potentials have
received considerable attention in past years \cite{SebbyStrabley06}.  In \cite{Milburn97}, the dynamics are discussed
both on a mean-field level, where the non-linear Schrödinger equation becomes the discrete self-trapping equation
\cite{Eilbeck84}, and in a quantum mechanically exact way. For large enough particle numbers, the two-site Hubbard
Hamiltonian can be approximated by the Josephson Hamiltonian \cite{Menotti01}, which is the reason for these systems to
be called Bose Josephson Junction (BJJ). A BEC in a double-well potential defines a representation of the rotation group
and hence a pseudo-spin which can be utilized for quantum information processing and studies of decoherence
\cite{Ferrini08}.  Other applications exploit the regime of weak tunneling, where the system behaves similarly to a
quantum nanostructure in the sequential tunneling limit \cite{Averin08,Cheinet08}.

A related strand of research is constituted by the investigation of Bose--Fermi mixtures. These have been studied mostly
on the mean-field level by solving non-linear Schrödinger equations for one \cite{Salerno05} or two wells
\cite{Caballero-Benitez09} or by composite-fermion methods on an optical lattice
\cite{Fehrmann04,Lewenstein04,Mering08}. Phase-diagrams have been computed for Bose--Fermi mixtures in three
\cite{Viverit00} or one dimensions \cite{Pollet06,Zujev08,Marchetti09}.

In this paper, we want to study the dynamics of a Bose--Einstein condensate in a double-well potential when a single
fermionic impurity is added to the system. The main question that comes into mind refers to the tunneling properties of
the fermion when the double-well is tilted: Does the BEC leave the fermion's tunnel properties unaffected, or does the
BEC expel the fermion to the other well against the potential gradient? Obviously, the answer depends on the relative
interactions between the two species. In our work, we consider the ground-state properties in the weak-tunneling limit.
We discuss the different regimes in the parameter space defined by the particle interactions and the implications of
large repulsion between the two species on the adiabatic and quasi-adiabatic dynamics. 

In Section~\ref{sec:Model}, we introduce the Hamiltonian and its basic properties. Then, in Section~\ref{sec:Phase}, we
compute the expectation value of the relative number operators indicating that the ground state shows different phases
defined by the repulsive forces. In particular, we find an avoided crossing between two states which are not connected
directly by the tunneling Hamiltonian. We calculate the splitting by an application of Brillouin--Wigner perturbation
theory in Section~\ref{sec:Spin}. In the last section of the paper, we consider time-dependent dynamics and show that by
using the avoided crossings due to the tunneling of the fermion as ``beam splitter'', we can construct a Mach--Zehnder
interferometer that allows us to transfer all population from one well into the other on a time scale that is only
defined by the tunneling properties of the fermion.

%% >>>1

\section{Model}\label{sec:Model} %% <<<1

The starting point for our discussion is the standard Bose Josephson Junction \cite{Milburn97}: a Bose--Einstein
condensate is loaded into an optical double-well potential such that only the ground states of each well are occupied.
Defining the relative number operator $n_\text{B} := \frac{1}{2}(n_\text{R} - n_\text{L})$ as the difference between the
number of bosons in the right and the left well, the Hamiltonian is, up to a constant depending on the total number of
particles $N = n_\text{R} + n_\text{L}$,~\cite{Averin08}
\begin{equation}
  H_\text{B} = -2\varepsilon n_\text{B} + U_\text{B} n_\text{B}^2 - \Delta_\text{B}\left(b_\text{L}^\dagger b_\text{R} + \text{h.c.}
\right).
\end{equation}
The double-well potential can be tilted generating an energy difference $2\varepsilon$ between the two wells' ground
states. The inter-particle repulsion of the bosons is of strength $U_\text{B}$, and tunneling between the two wells
occurs with amplitude $\Delta_\text{B}$. The operators $b_i$, $b_i^\dagger$, annihilate and create a boson in the
respective well. The two most prominent regimes of the Bose Josephson Junction are the superfluid-like regime
$U_\text{B} \ll \Delta_\text{B}$, where the particles are delocalized over the two wells, and the Mott-like regime
$\Delta_\text{B} \ll U_\text{B}$, where the particles are localized. In the latter case, tunneling is considered 
a perturbation to the Hamiltonian, which is diagonal in the relative number-state representation. Tilting the double-well
potential by adjusting $\varepsilon$, the bosons will tunnel into the other well one by one leading to a staircase
profile for the expectation value $\erw{n_\text{B}}(\varepsilon)$ \cite{Averin08,Cheinet08}.

We now consider an additional single fermion, or, since in that case the particle statistics do not matter at all, an
atom of a different species than the constituents of the condensate. Its dynamics are governed by the very same
Hamiltonian without the repulsive term
\begin{equation}
  H_\text{F} = -2\varepsilon n_\text{F} - \Delta_\text{F}\left(c_\text{L}^\dagger c_\text{R} + \text{h.c.} \right),
\end{equation}
with appropriately labeled constants and $c_i$, $c_i^\dagger$ being the fermion's annihilation and creation operators.
We assume that the mutual interaction of both species is proportional to $\sum_{\alpha = \text{L,R}}n_\text{B}^\alpha
n_\text{F}^\alpha$ \cite{Averin08}, which in the relative number representation reads
\begin{equation}
  H_\text{B--F} = 2U_\text{B--F} n_\text{B} n_\text{F}
\end{equation}
plus a constant.  The full Hamiltonian of our system is thus
\begin{equation}\label{eq:H}
  H = H_\text{B} + H_\text{F} + H_\text{B--F}.
\end{equation}
The double-well potential in the two-mode approximation defines a representation of the rotation group $SU(2)$, hence a
pseudo-spin on the Bloch-sphere whose length is the number of particles. The $z$-direction of this spin encodes the
position information of the particles and is given by the relative number operator $n_\text{B/F}$ with eigenvalues
$m_\text{B/F}$ \cite{Milburn97}.

%% >>>1

\section{Phase Diagram}\label{sec:Phase} %% <<<1

As it is shown in \cite{Averin08,Cheinet08}, by adjusting $\varepsilon$ adiabatically and thereby tilting the double-well
potential, the Bose Josephson Junction shows single-particle tunneling and a staircase profile of the expectation
value of the relative number operator. In our case, where an additional species, although only a single particle of it, is
present in the system, we expect the same behavior in the tunneling regime $U_\text{B}, U_\text{B--F} \gg \Delta_\text{B/F}$: tilting the potential
makes the particles tunnel from one well to the other. Since there are two species of atoms present in the potential,
the obvious question to ask is, which species will tunnel? 

In general, transitions will be shown to be interaction-mediated, that is the relative magnitude of the repulsive
interactions will determine the tunneling species. If there are no interactions as for instance in the case $N = 1$, or
at the exact threshold where the tunneling species changes, $U_\text{B} = U_\text{B--F}$, the profile of the transitions
in $\erw{n_\text{B/F}}$ is heavily influenced by the kinetic energy.

\begin{center}
  \begin{figure}
    \includegraphics[width=\linewidth]{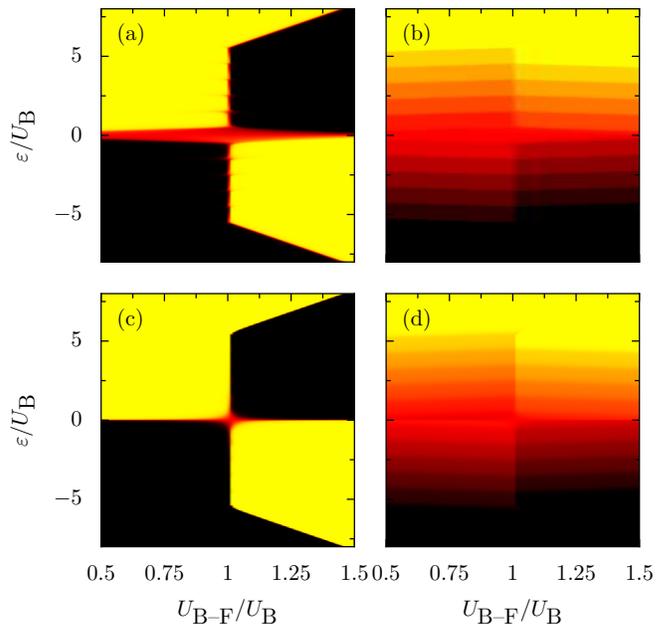}
    \caption{(Color online) Number expectation for a single fermion and $N_\text{bosons} = 11$. Left: $\erw{n_\text{F}}$, right:
    $\erw{n_\text{B}}$. The parameters are $\Delta_\text{F} = 0.01$ and $\Delta_\text{B} = 0.05$ in the upper plots and
    exchanged for the lower plots.  The color code is Black--Red--Yellow for Negative--Zero--Positive.
    \label{fig:Mixture}}
  \end{figure}
\end{center}

In Fig.~\ref{fig:Mixture}, we show numerical results for $\erw{n_\text{B/F}}$ as a function of inter-species interaction
$U_\text{B--F}$ and tilt $\varepsilon$ for two pairs of values of $\Delta_i$.  The most obvious characteristic is the
change of the tunneling species at the first resonance for $U_\text{B--F} = U_\text{B}$.  Close to this point, the fermionic
expectation values show traces of attempted tunneling (a). For larger inter-species interaction, the bosonic expectation
values are shifted by one due to the presence of the fermion in the well with higher energy (b,d). For larger fermionic
tunnel amplitude, the bosons show negative compressibility $\kappa_\text{B} = \mathrm{d}\erw{n_\text{B}}/\mathrm{d}\varepsilon$ close to
zero tilt (d). In the following, we shall discuss and explain these phenomena in detail.

Since we are in the tunneling regime with well separated resonances, we begin by restricting the bosonic Hilbert space
at each resonance to the subspace spanned by the eigenstates of $n_\text{B}\ket{m} = m\ket{m}$, $\ket{m}$, $\ket{m +
1}$. The fermionic Hilbert space is also two dimensional, such that we have
\begin{subequations}
  \begin{align}
    H_\text{B} & = \begin{pmatrix}
      U_\text{B} m^2 - 2\varepsilon m & -\lambda_m \\
      -\lambda_m & U_\text{B}(m+1)^2 - 2\varepsilon(m+1)
    \end{pmatrix},\\
    H_\text{F} & = \begin{pmatrix}
      \varepsilon & - \Delta_\text{F}\\
      -\Delta_\text{F} & -\varepsilon
    \end{pmatrix},\\
    H_\text{B--F} & = 2 U_\text{B--F} n_\text{B} n_\text{F}
  \end{align}
\end{subequations}
in the respective basis. In the same basis, the relative number operators take the form
\begin{equation}
  n_\text{B} = \begin{pmatrix}
    m & 0 \\
    0 & m+1
  \end{pmatrix},
\qquad
n_\text{F} = \begin{pmatrix}
  -\frac{1}{2} & 0\\
  0 & \frac{1}{2}
  \end{pmatrix}.
\end{equation}
The off-diagonal elements of $H_\text{B}$ are the tunnel amplitudes times the matrix element of the bosonic operators
$b_\alpha$, $b_\alpha^\dagger$, $\lambda_m := \Delta_\text{B}\sqrt{(\frac{1}{2}N_\text{B} + m + 1)(\frac{1}{2}N_\text{B} -
m)}$ \cite{Averin08}.
In the absence of tunneling, the Hamiltonian is already diagonal with energies
\begin{equation}  
E_{\ket{m;\pm\frac{1}{2}}} = U_\text{B}m^2 \pm U_\text{B--F}m - 2\varepsilon\left(m\pm\frac{1}{2}\right).
\end{equation}
If we place the system at a point, where tunneling of a particle should occur, we have either the states $\ket{m;\pm
\frac{1}{2}}$ and $\ket{m+1;\pm\frac{1}{2}}$ degenerate or $\ket{m;\pm\frac{1}{2}}$ for a bosonic or a fermionic
resonance, respectively. Equating the unperturbed energies of the Hamiltonian, we find that the fermion can tunnel for
$\varepsilon = m U_\text{B--F}$, and the boson can tunnel for $\varepsilon = U_\text{B}(m + \frac{1}{2}) \pm
\frac{1}{2}U_\text{B--F}$ depending on the position of the fermion. In particular, if we consider two values of the
boson--fermion interaction, $U_{\text{B--F},\pm}$ with $\frac{1}{2}(U_{\text{B--F},+} + U_{\text{B--F},-}) =
U_\text{B}$,  at $U_\text{B--F} = U_{\text{B--F},+}$, the state $\ket{m; -\frac{1}{2}}$ has the same energy as
$\ket{m-1; \frac{1}{2}}$ has at $U_\text{B--F} = U_{\text{B--F},-}$ This property is seen in Figs.~\ref{fig:Mixture}(b)
and (d), where the bosonic resonances at equal distances left and right from $U_\text{B--F} = U_\text{B}$ differ by one
boson exactly.

Approaching the first resonance of the system from large negative $\varepsilon$, the ground state is $\ket{m = -N/2;
-\frac{1}{2}}$; if the fermion tunnels first, we have to equate its energy with $E_{-N/2;1/2}$ otherwise with
$E_{-N/2+1;-1/2}$. From the above formula for the energy, we see that the difference between the positions of these
resonances is $\varepsilon_\text{F} - \varepsilon_\text{B} = \frac{1}{2}\left[(1-N)(U_\text{B--F} - U_\text{B})
\right]$. Hence if $U_\text{B--F} > U_\text{B}$, the fermion will tunnel first and a boson otherwise. Repeating this
calculation for arbitrary $m$ shows that this condition is independent of the resonance in question: for large enough
inter-species repulsion, the fermion will tunnel first, otherwise last.

This result is intuitively clear, as the condition states that for $U_\text{B--F} > U_\text{B}$, keeping the fermion
together with the bosons costs more energy than keeping the additional boson. Hence the fermion will be expelled to the
other well. Noting that such behavior occurs for finite $\varepsilon$, the fermion will move to the potential well
with higher energy, hence against the potential gradient.

Directly at the degenerate point $U_\text{B--F} = U_\text{B}$, it cannot be decided from the atomic interactions alone,
which particle will tunnel first. In this situation, we have to include the tunnel Hamiltonian, as now the kinetic
energy of the particles will decide. At $\varepsilon = m U_\text{B} = mU_\text{B--F}$, the unperturbed Hamiltonian's
ground-state is four-fold degenerate:
\[
\ket{m-1;\frac{1}{2}}, \ket{m; -\frac{1}{2}}, \ket{m;\frac{1}{2}}, \ket{m+1; -\frac{1}{2}}
\]
all have the same energy. Including the tunneling the Hamiltonian in this basis reads
\begin{equation}
  \tilde{H} = \begin{pmatrix}
    0 & 0 & -\lambda_{m-1}& 0\\
    0 & 0 &  -\Delta_\text{F} & -\lambda_m\\
    -\lambda_{m-1} & -\Delta_\text{F} & 0 & 0\\
    0 & -\lambda_m & 0 & 0
  \end{pmatrix}
\end{equation}
plus a constant. This Hamiltonian has a bi-quadratic characteristic polynomial, which can be solved explicitly, leading to the ground-state
expectation values
\begin{subequations}
  \begin{align}
    \erw{n_\text{F}} & = \frac{\Delta_\text{B}^2 m}{\sqrt{\Delta_\text{F}^4 + 4\Delta_\text{B}^4 m^2 + \Delta_\text{B}^2\Delta_\text{F}^2\left[-4m^2 + N(N+2) \right]}}\label{eq:Res1}\\
    \erw{n_\text{B}} & = m - \erw{n_\text{F}}.\label{eq:Res2}
  \end{align}
\end{subequations}
For the first resonance from positive/negative $\varepsilon$, the state $\ket{m\pm 1;\mp \frac{1}{2}}$ drops out, and
the expression simplifies to
\begin{subequations}
\begin{align}
  \erw{n_\text{F}} & = \pm\frac{1}{2}\frac{\Delta_\text{B}^2 N}{\Delta_\text{B}^2 N + \Delta_\text{F}^2}\\
  \erw{n_\text{B}} & = \pm\frac{N}{2}\left(1 - \frac{\Delta_\text{B}^2}{\Delta_\text{F}^2 + \Delta_\text{B}^2 N} \right).
\end{align}
\end{subequations}
Note that these are discrete values evaluated at $\varepsilon = mU_\text{B--F}$ and are \textit{not} continuous in $m$.

From these expressions, we directly infer that at the first resonance, a fast tunneling species, i.e., one with large
$\Delta_i$, will hamper the other and restrict its number expectation value to the asymptotic value of $\pm \frac{1}{2}$
or $\pm N/2$, respectively.

\begin{center}
  \begin{figure}
    \begin{minipage}[c]{\linewidth}
      \includegraphics[width=\linewidth]{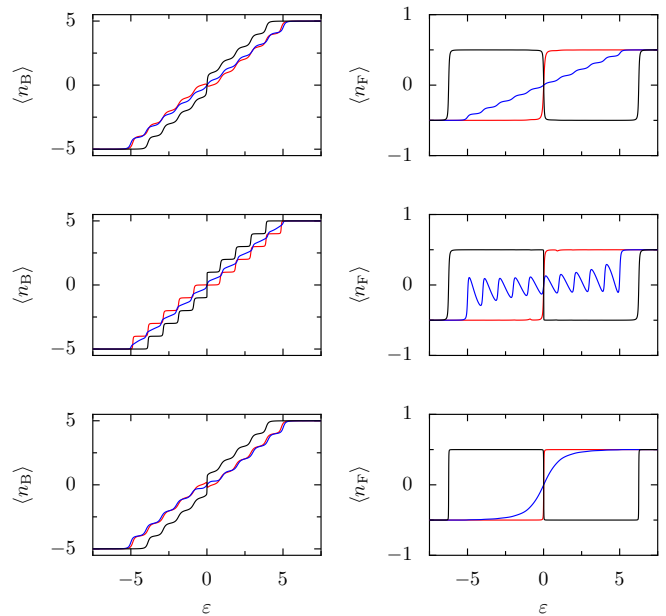}
    \end{minipage}
    \caption{(Color online) Occupation number difference $\erw{n_i}$ (bosons left/fermions right) for a mixture of 10 bosons and one
    fermion. The plots are for different values of the Bose--Fermi repulsion $U_{\text{B--F}}\in\{0.75, 1,1.25\}$ in
    red/blue/black. The tunneling amplitudes are $\Delta_\text{B} = \Delta_\text{F} = 0.05$ (top), $\Delta_\text{B} =
    0.01, \Delta_\text{F} = 0.05$ (middle), and $\Delta_\text{B} = 0.05$, $\Delta_\text{F} = 0.01$ (bottom).}\label{fig:slices}
  \end{figure}
\end{center}

By considering the limit of large and small tunneling amplitudes, we can significantly simplify the
expressions~(\ref{eq:Res1}, \ref{eq:Res2}). Also, as for $U_\text{B--F} < U_\text{B}$ only single-particle processes
are present, the situation at $U_\text{B--F} = U_\text{B}$ can be used as a good approximation for the ground-state
properties in that regime. For $U_\text{B--F} > U_\text{B}$ we observe processes where a larger number of particles tunnels
simultaneously. This will be discussed in Section~\ref{sec:Spin}. In the present case, the asymptotic expectation values
at the resonances are
\[
\erw{n_\text{F}} \to \begin{cases}
  0 & \text{for }\Delta_\text{B}/\Delta_\text{F} \to 0\\
  \frac{m}{\sqrt{1+N(N+2)}} & \text{for } \Delta_\text{B} = \Delta_\text{F}\\
  \frac{1}{2}\sgn(m) & \text{for } \Delta_\text{B}/\Delta_\text{F} \to \infty
\end{cases}
\]
and
\[
\erw{n_\text{B}} \to \begin{cases}
  m & \text{for }\Delta_\text{B}/\Delta_\text{F} \to 0\\
  m\left(1 - \frac{1}{\sqrt{1+N(N+2)}} \right) & \text{for } \Delta_\text{B} = \Delta_\text{F}\\
  m - \frac{1}{2}\sgn(m) & \text{for }\Delta_\text{B}/\Delta_\text{F} \to \infty.
\end{cases}
\]
For $\Delta_\text{B} = \Delta_\text{F}$, shown in Fig.~\ref{fig:slices}~(top panels), the fermionic expectation value shows steps linear in the
resonance number $m$.  For $\Delta_\text{B} < \Delta_\text{F}$, Fig.~\ref{fig:slices}~(middle panels), the fermionic expectation shows oscillations
about $\erw{n_\text{F}} = 0$.  In the other case, $\Delta_\text{B} > \Delta_\text{F}$, shown in Fig.~\ref{fig:slices}~(lower panels), we see that the
presence of the fermion and in particular its tunnel amplitude do influence the bosons in such a way that for
$\Delta_\text{B} \gg \Delta_\text{F}$, i.e., a strongly localized fermion, the relative number expectation value
approaches $m - \frac{1}{2}\sgn(m)$ which is the same as in the case $U_\text{B--F} < U_\text{B}$. Also the
expectation-value of the fermionic relative number operator approximates a step function.

\begin{center}
  \begin{figure}
    \includegraphics[width=.9\linewidth]{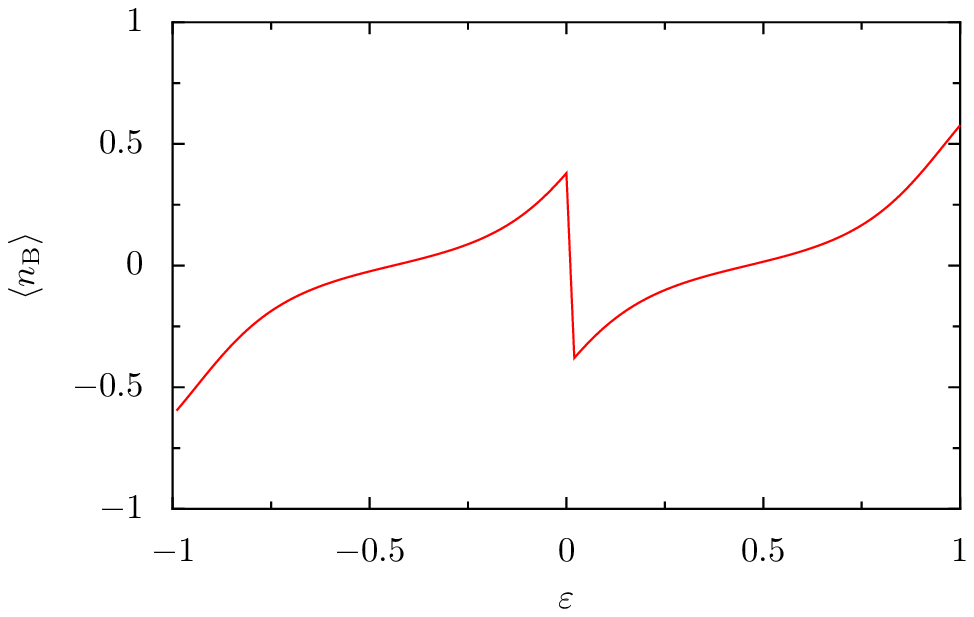}
    \caption{(Color online) Negative compressibility $\kappa_\text{B} = \mathrm{d}\erw{n_\text{B}}/\mathrm{d}\varepsilon$ due to low fermionic tunnel amplitude $\Delta_\text{F} = 10^{-4}$, $\Delta_\text{B}
    = 0.05$, and $U_\text{B--F}/U_\text{B} = 0.9$.\label{fig:negcomp}}
  \end{figure}
\end{center}

In the case of low fermionic tunnel amplitude, the resonance at zero tilt $\varepsilon = 0$, with $m = 0$ will also show
negative compressibility $\kappa_\text{B} = \mathrm{d}\erw{n_\text{B}}/\mathrm{d}\varepsilon  < 0$, because in the limit $\Delta_\text{F} = 0$, $\lim_{m\nearrow 0}
\erw{n_\text{B}} = \frac{1}{2}$ and $\lim_{m\searrow 0}\erw{n_\text{B}} = -\frac{1}{2}$, see Fig.~\ref{fig:negcomp}. A
quantitative result is given in \cite{Averin08} for $0 \le \varepsilon \ll U_\text{B--F} < U_\text{B}$.

%% >>>1

\section{Zero-bias Spin Flip}\label{sec:Spin} %% <<<1

Due to the symmetries of the Hamiltonian without tunneling, $\Delta_\text{B/F} = 0$, at $\varepsilon = 0$, the states
$\ket{m_\text{B};m_\text{F}}$ and $\ket{-m_\text{B};-m_\text{F}}$ are degenerate. In particular, this holds for the
ground state of the non-interacting system. In general such a degeneracy for the ground state always occurs at a
resonance.  In our system, however, the degenerate states do not only differ by a single particle having changed its
position, but rather correspond to the exchange of positions for two species and, depending on the inter-species
repulsion $U_\text{B--F}$, a larger number of bosons. Since, as we have mentioned above, the relative number operators
are equivalent to the $z$-components of the pseudo-spin defined by the double-well potential, this transition
corresponds to a flip of the combined pseudo-spin of bosons and fermion.

\subsection{Higher-order degenerate perturbation theory}\label{sec:BW} %% <<<2

In contrast to the single-particle resonance, whose physics is that of an avoided crossing and degenerate perturbation
theory, the zero-bias spin flip is not easily amenable to degenerate perturbation theory, because the involved states
are not directly connected by the tunneling term in the Hamiltonian. Although a numerical treatment of the problem is
straightforwardly implemented, perturbation theory allows for a simple analytical approach to the resonance and provides a
very good approximation for the lineshape.

For general considerations, let $\ket{m}$ and $\ket{-m}$ be eigenstates of the unperturbed Hamiltonian with energies
$E_{\pm m} = \varepsilon_0 \pm\epsilon$ such that they are degenerate for $\epsilon = 0$.  Let $\xi V$ be a
perturbation, with a scalar $\xi$ to keep track of orders of magnitude, and $\fbkt{-m}{V}{m} = 0$.  Assume the shortest
chain of matrix elements of $V$ that connects the two degenerate states via intermediate states $\ket{n_j}$ has length
$k$. Brillouin--Wigner perturbation theory to lowest order in $\xi$ yields for the splitting \cite{Garanin91}
\begin{multline}\label{eq:DeltaE}
  \Delta E_m = 2 \xi^k V_{m,n_1} \frac{1}{E_{n_1} - E_m}V_{n_1, n_2}\frac{1}{E_{n_2} -E_{m}}\cdots\\ 
  \cdots \frac{1}{E_{n_k} - E_m}V_{n_k,-m}
\end{multline}

The lineshape of the expectation value of the relative occupation number $n = m\bigl(\proj{m} - \proj{-m}\bigr)$
restricted to the degenerate subspace as a function of $\epsilon$ is thus to lowest order in $\xi$,
$\erw{n}(\epsilon) \approx \abs{m}\epsilon/\sqrt{\frac{1}{4}\Delta E_m^2 + \epsilon^2}$.

%% >>>2

\subsection{Application to the Bose--Fermi mixture} %% <<<2

Let us apply this reasoning to the spin flip at zero bias of a Bose--Fermi mixture in a double-well potential.  In the
regime with $U_\text{B--F} > U_\text{B}$, the spin-flip transition is $\ket{-m;\frac{1}{2}}\mapsto \ket{m;
-\frac{1}{2}}$. To construct the chain of intermediate states connecting the two degenerate states by
single-particle processes, we climb up the angular-momentum ladder in $m$. At a certain $m = m_0$, the fermion
jump is included. Then, due to the different repulsion energies, the energy-denominators are different and the chain
naturally splits into two products. Consider the energy differences
  \begin{subequations}
    \begin{align}
      G^{-1}_<(n) := & E_{n,\frac{1}{2}} - E_{-m,\frac{1}{2}}\nonumber \\ = & (n^2 - m^2)U_\text{B} + (n+m)U_\text{B--F}\\
      G^{-1}_>(n) := & E_{n,-\frac{1}{2}} - E_{-m,\frac{1}{2}}\nonumber \\ = & (n^2 - m^2)U_\text{B} - (n-m)U_\text{B--F}
    \end{align}
  \end{subequations}
  for bosonic transitions left ($G^{-1}_<$) and right ($G^{-1}_>$) of $m_0$. The fermionic jump is given by $G^{-1}_>(m_0)$,
  which leads to the result
  \begin{widetext}
  \begin{equation}
    \Delta E = 2\Delta_\text{F}\left(\prod_{k = -m}^m \lambda_k\right) \sum_{m_0 = -m}^m\left[ \prod_{n = m+1}^{m_0}
    G_<(n) \prod_{n = m_0}^{m-1} G_>(n)\right].
  \end{equation}
\end{widetext}
The fermionic jump is included as the first factor of the last product. Here, we use the convention that
a product without factors, eg., $\prod_m^n$ with $n< m$, is unity.

\begin{center}
  \begin{figure}
    \begin{minipage}[c]{\linewidth}
      \includegraphics[width=\linewidth]{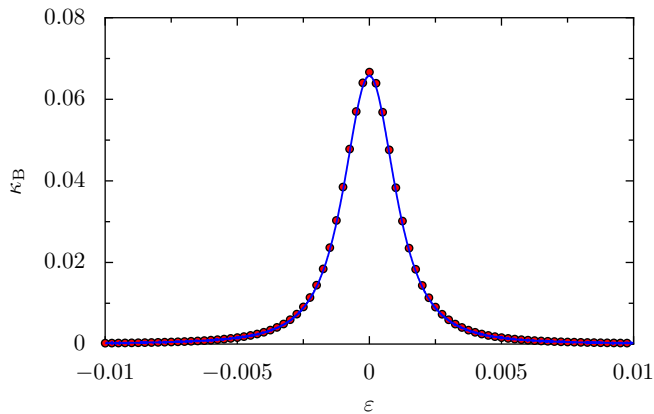}
    \end{minipage}
    \caption{(Color online) The Brillouin--Wigner result (dots) and the numerically obtained lineshape (solid line) for the
    compressibility $\kappa_\text{B} = \text{d}\erw{n_\text{B}}/\text{d}\varepsilon$ of the zero-bias spin flip.  Note
    that the energy scales have to be separated by about two orders of magnitude $J/U \sim 10^{-2}$. Here $U_\text{B} =
    1$, $U_\text{B--F} = 2$ , $\Delta_\text{B} = 0.1$, and $\Delta_\text{F} = 0.05$, such that $\Delta E = 6
    \Delta_\text{B}^2 \Delta_\text{F}$.\label{fig:spin-flip-fit}}
  \end{figure}
\end{center}

As an example, we consider the smallest non-trivial mixture: $N = 2$. For $U_\text{B--F} > U_\text{B}$, the
spin-flip transition at zero bias is $\ket{-1;1/2}\mapsto\ket{1;-1/2}$, which are connected by three paths
\begin{equation}
  \ket{-1;\frac{1}{2}} \mapsto 
  \begin{Bmatrix} 
    \ket{0;\frac{1}{2}} & \mapsto & \ket{1;\frac{1}{2}}\\[4pt]
    \ket{0;\frac{1}{2}} & \mapsto & \ket{0; -\frac{1}{2}}\\[4pt]
    \ket{-1;-\frac{1}{2}} & \mapsto & \ket{0; -\frac{1}{2}}
  \end{Bmatrix} 
  \mapsto \ket{1; -\frac{1}{2}}.
\end{equation}
The level-splitting is thus
\begin{equation}
   \Delta E = 4 \Delta_\text{B}^2 \Delta_\text{F} \frac{1}{U_\text{B} - U_\text{B--F}}\left[\frac{-1}{U_\text{B--F}} + \frac{1}{U_\text{B}
  - U_\text{B--F}} \right]
\end{equation}
and, since $m = 1$, as a function of the tilt $\varepsilon$, $\erw{n_\text{B}}(\varepsilon) = \varepsilon/\sqrt{\varepsilon^2 +
\frac{1}{4}\Delta E^2}$. In Fig.~\ref{fig:spin-flip-fit}, we show the numerical data for $\kappa_\text{B} =
\mathrm{d}\erw{n_\text{B}}/\mathrm{d}\varepsilon$ as well as the line shape computed with Brillouin--Wigner perturbation
theory.

Of course, the result is not absolutely accurate, as in the tails of the resonance, we do \textit{not} recover the
asymptotic states with well-defined occupation numbers, but rather the exact eigenstates of the full Hamiltonian. If we
look at the numbers of Fig.~\ref{fig:spin-flip-fit}, Brillouin--Wigner theory gives $\Delta E = 3\cdot 10^{-3}$, whereas
the numerically evaluated splitting is approximately $2.88\cdot 10^{-3}$.

%% >>>2
%% >>>1

\section{Landau--Zener Dynamics}\label{sec:LZ} %% <<<1

\begin{center}
  \begin{figure*}
    \begin{minipage}[c]{\linewidth}
      \includegraphics[width = \linewidth]{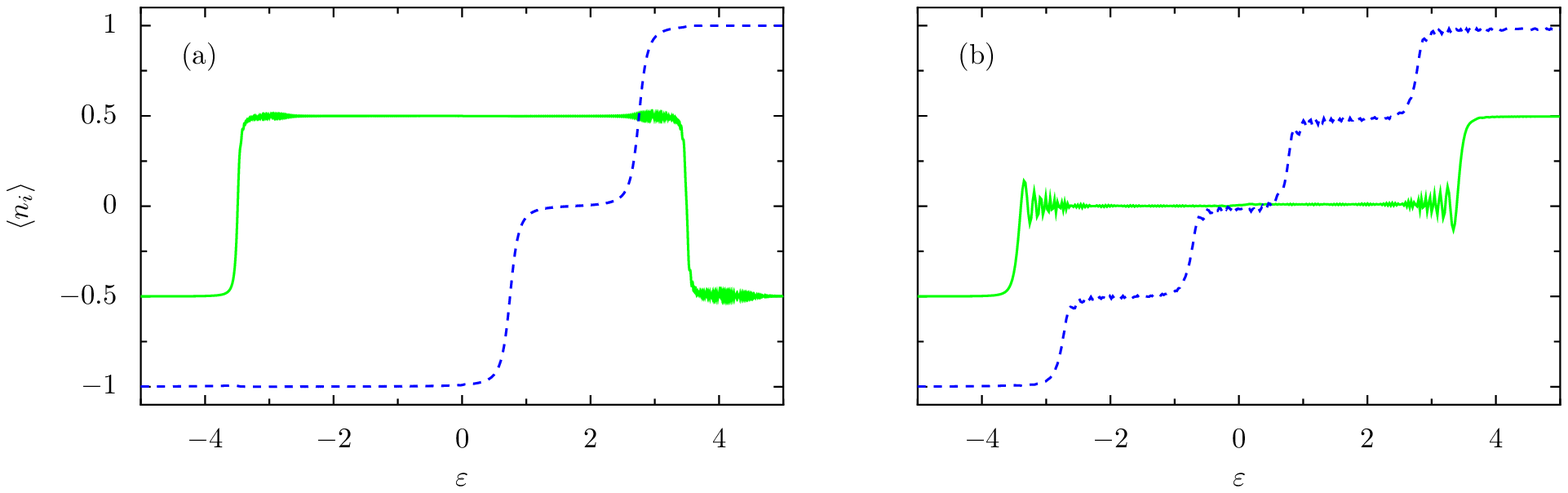}
    \end{minipage}
    \caption{(Color online) Expectation values $\erw{n_i}$ of the relative number operator for fermions (green/solid) and bosons
    (blue/dashed) in a numerical Landau--Zener experiment on the Bose--Fermi mixture. $\Delta_\text{B} = 0.1$,
    $\Delta_\text{F} = 0.05$, $U_\text{B} = 2$, $U_\text{B--F} = 3.5$, (a) $\alpha = 0.001$, (b) $\alpha \approx
    0.011$.\label{fig:AD_MZ}}
  \end{figure*}
\end{center}

\begin{center}
  \begin{figure}
    \begin{minipage}[c]{\linewidth}
      \includegraphics[width = \linewidth]{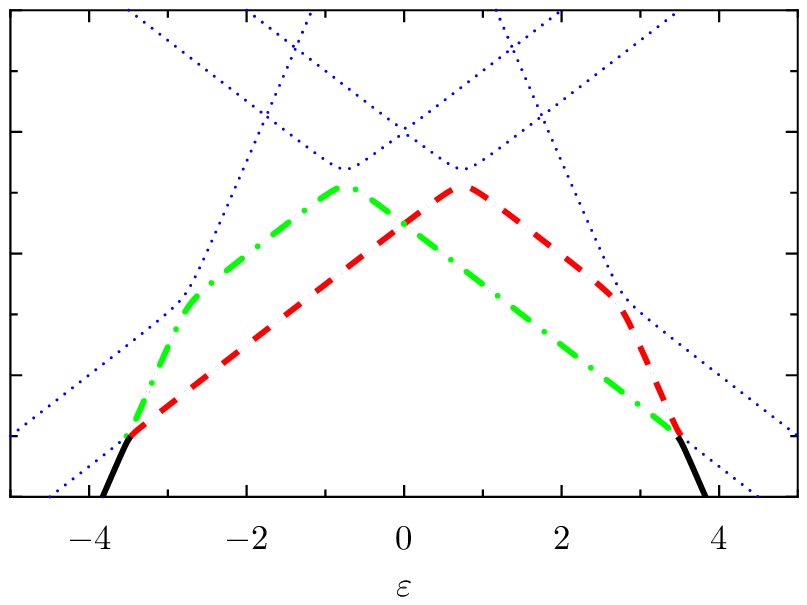}\hfill
    \end{minipage}
    \caption{(Color online) Spectrum of the Bose--Fermi mixture and the path that is traversed by a linear scan of
    $\varepsilon$. The branches that define the two equivalent arms of the Mach--Zender interference are shown in
    red/dashed and green/dash-dotted, black/solid denotes the in- and outgoing beam.\label{fig:spectrum}}
  \end{figure}
\end{center}

In the previous sections, we have focused on the adiabatic dynamics of population transfer from one well into the other
by increasing the tilt $\varepsilon$ of the potential. In a realistic scenario, however, we will always adjust the tilt
within finite time, hence with finite velocity $\mathrm{d}\varepsilon/\mathrm{d} t$. This means that we have to take
into account Landau--Zener physics of quasi-adiabatic transitions \cite{Zener32, Garanin03, Garanin04, Petta10,
Oliver05}. Depending on the velocity and the splitting of the states at a resonance, the population transfer is thus
heavily influenced...

In the regime $U_\text{B--F} > U_\text{B}$ of our model, the zero-bias spin-flip resonance is so narrow that it could be
very difficult to traverse it adiabatically as the necessary velocities, depending on the splitting, are very small. In
\cite{Schlagheck10}, such an effect is also seen and successfully exploited to achieve the population transfer sought
for. Let us choose a velocity $\alpha$ for the change of $\varepsilon (t) = \alpha t$ in time that allows to pass the
single-particle resonances adiabatically. Starting at infinite negative time, the fermion tunnels from left to right at
finite negative $\varepsilon$ as it is expected for $U_\text{B--F} > U_\text{B}$. The spin-flip resonance is, however,
passed completely diabatically, such that all particles stay where they are and do not exchange places. At this point
the system is no longer in its ground state but in the first excited state. Further tilting of the potential causes the
bosons to tunnel to the lower-lying well step by step. Finally, the system adiabatically passes an avoided crossing of
two excited states, where again, the fermion changes sides and tunnels back into the \textit{left}, i.e., higher-lying
well, as is seen in Fig.~\ref{fig:AD_MZ}(a). There it will stay until other processes than those describe by our
Hamiltonian cause it to tunnel again. The diabatic traversing of the spin-flip resonance thus causes the transition of
the system from its ground state at $t\to -\infty$ to the first excited state at $t\to \infty$. Lowering the velocity
allows us to pass to the adiabatic regime also for the central resonance, but the more bosons we have in the condensate,
the slower we would have to adjust $\varepsilon(t)$.

In order to facilitate adiabatic population transfer from one well to the other, we make use of a constructive
interference effect. The spectrum of the Bose--Fermi mixture, as it is shown in Fig.~\ref{fig:spectrum} is symmetric
about $\varepsilon = 0$.  If we adjust the tunnel amplitudes such that $\Delta_\text{F} < \Delta_\text{B}$, we can
control the Landau--Zener physics of the fermionic jump and still pass all other bosonic resonances almost
adiabatically. This allows us to restrict the problem to only consider the ground state $\ket{g}$ and the first excited
state $\ket{e}$ of the Hamiltonian as depicted in Fig.~\ref{fig:spectrum} for the case of $N = 3$. Passing the first
avoided crossing by adjusting the tilt $\varepsilon(t)$ amounts to a unitary transformation of the asymptotic states
\[
U = \begin{pmatrix}
  \cos \Theta & \sin \Theta\\
  -\sin \Theta & \cos \Theta
\end{pmatrix}
\]
with the angle $\Theta$ depending on the velocity $\alpha$ and the splitting $\Delta E$ of the two states $\cos^2 \Theta
= 1- \exp\left(-\frac{2\pi}{\hbar} \frac{\Delta E^2}{\alpha} \right)$ \cite{Zener32}.  The
spin-flip transition is very narrow, such that we traverse it diabatically, hence exchanging ground and excited state,
which corresponds to the unitary transformation $\sigma_x$. The second avoided crossing is identical to the first; however, it
is passed in opposite direction, whence the invoked unitary is $U^\dagger$. Since both arms of the interferometer are
identical --- they can be mapped by $\varepsilon \mapsto -\varepsilon$ onto each other, the accumulated dynamical phase
is the same and amounts to a global phase factor, which is of no importance for our purpose. 

With these assumptions, the constructed Mach--Zehnder interferometer is the map 
\[
U\sigma_x U^\dagger = \sigma_x \cos(2\Theta) - \sigma_z \sin(2\Theta)
\]
For $\Theta = \pi/4$, $U\sigma_x U^\dagger = -\sigma_z$: the ground state of the full Hamiltonian will again be mapped
onto the ground state and population can be transfered completely. Considering the
avoided crossings as a beam-splitter for an incoming ground-state wave function \cite{Petta10}, this choice of $\Theta$ amounts
to a velocity $\alpha$ for which the Landau--Zener transition splits the wave-function exactly in two halves $\ket{g} \mapsto
\frac{1}{\sqrt{2}}\bigl(\ket{g} + \ket{e}\bigr)$.

In the numerical data in Fig.~\ref{fig:AD_MZ}(b), the achieved population transfer is almost perfect.  The expectation
values of the relative number operators show some oscillations in the fermionic part, which are due to the coherent
superposition created by the Landau--Zener transition. The remainder of the profile is, however, well in accordance with
the adiabatic picture for atom counting, only that, since the wave function is split by the avoided crossing, we do not
count full but half atoms, as the change in $\erw{n_i}$ is $1/2$ instead of the expected $1$.

The important advantage of this approach over a completely adiabatic transfer is that with this interferometer, we only
need to adjust the Landau--Zener transition of the single fermionic resonance. This, however, is independent of the
number of bosons in the system, and we can use a much higher velocity than in the purely adiabatic regime where the
Landau--Zener physics of the spin-flip transition are taken fully into account.

%% >>>1

\section{Conclusion} %% <<<

In this paper, we have investigated a Bose--Fermi mixture in a double-well potential with the restriction that while the
number of bosons is arbitrary, the number of fermions is fixed at one. We thus discussed the influence of a single
fermionic impurity on the ground-state properties of the Bose Josephson Junction, when the potential was allowed to be
tilted. It has turned out that rather than the bosons, it is the fermion that is affected most by the boson--fermion
interactions. We have separated the dynamics into two regimes: one, where the bosons and the fermion live side-by-side,
and one, where the fermion is expelled from the BEC towards the higher-lying well. In this regime, we have found a
zero-bias spin-flip, where the fermion and a large number of bosons change places. The physics of this process, in
particular the level splitting and the lineshape of the observed resonance in the relative particle number, are well
accessible by Brillouin--Wigner perturbation theory.  Since many particles are involved in this transition, complete
population transfer of all particles from the left to the right well upon tilting the potential is hardly possible
anymore, when we assume the realistic case of slow but not infinitely slow adjustment of the potential tilt
$\varepsilon$. Instead, we have shown, how the fermionic resonance in the regime $U_\text{B--F} > U_\text{B}$ can be
employed as a beam-splitter of a Mach--Zehnder interferometer to achieve population transfer at much higher tilting
speeds than would be necessary to traverse the zero-bias spin-flip resonance adiabatically.

There are several direction in which we want to extend our investigation. In the case of very large $N$, the
Gross--Pitaevskii equation provides a much simpler description of the BEC than can be achieved by numerical
diagonalization of the two-site Bose--Hubbard Hamiltonian. We will therefore ask for the lowest-order corrections of the
dynamics of a Bose Josephson Junction with large $N$ in the presence of a single fermionic impurity. The trade-off for
the reduction to the Gross--Pitaevskii equation is, however, the extension to the study of nonlinear eigenvalue
problems.

The second direction refers to the number of fermions in the double-well potential. Experiments of two fermions with
opposite spin in such a system without a BEC have been conducted by Trotzky \textit{et al.} \cite{Trotzky08}. The
orbital wave function of two fermions with opposite spin in a double-well potential is decomposed into a singlet and
three triplet states with respect to the pseudo-spin defined by the double-well potential.  Interestingly, the singlet
does not couple to the bosons at all. On the contrary, the state $\ket{\uparrow, \downarrow}$, that is a configuration
with \textit{definite} position of the spin-up fermion in the left well and the spin-down particle in the right well, is
a superposition of the singlet and the $J_z = 0$ triplet, which, however, does couple to the BEC. In this configuration
we expect the time-dependent dynamics to show very interesting phenomena, which could be observed experimentally.

%% >>>

This work has been supported by the Swiss SNF and the NCCR Nanoscience.

%% >>>

%% \bibliography{../../../bibliography/Cold_Atoms,../../../bibliography/tunnelling,../../../bibliography/books,../../bibliography/articles,../../bibliography/math,../../bibliography/optics}

\providecommand{\SortNoop}[1]{}
 %% >>>

\end{document}